1　Does your gene need a background check? How genetic background impacts the

2　analysis of mutations, genes, and evolution






5　Author list: Christopher H. Chandler[1,2], Sudarshan Chari[1] and Ian Dworkin[1*]

6　* Author for correspondence, Email: idworkin@msu.edu



8　Author affiliation:

9　　1 - Department of Zoology, BEACON Center for the study of Evolution in Action.

10　Michigan State University.

11　　2 - Department of Biological Sciences, SUNY Oswego, Oswego, NY.










The premise of genetic analysis is that a causal link exists between phenotypic and allelic variation. Yet it has long been documented that mutant phenotypes are not a simple result of a single DNA lesion, but rather are due to interactions of the focal allele with other genes and the environment. Although an experimentally rigorous approach, focusing on individual mutations and isogenic control strains, has facilitated amazing progress within genetics and related fields, a glimpse back suggests that a vast complexity has been omitted from our current understanding of allelic effects. Armed with traditional genetic analyses and the foundational knowledge they have provided, we argue that the time and tools are ripe to return to the under-explored aspects of gene function and embrace the context-dependent nature of genetic effects. We assert that a broad understanding of genetic effects and the evolutionary dynamics of alleles requires identifying how mutational outcomes depend upon the "wild-type" genetic background. Furthermore, we discuss how best to exploit genetic background effects to broaden genetic research programs.



1   **What are genetic background effects?**

2        Although many traits vary phenotypically (and genetically) in natural populations,

3   some appear qualitatively similar across unrelated individuals, as long as those

4   individuals possess a "wild-type" genotype. This phenomenon is often depicted with

5   "genotype-phenotype maps", diagrams illustrating how similar phenotypes can be

6   produced in spite of variation in both genotypes and in underlying intermediate

7   phenotypes such as gene expression (Figure 1A). However, when particular mutations

8   (whether induced or natural variants) are placed into each of these different wild-type

9   backgrounds, the phenotypic consequences of that allele may be profoundly different

10   (Figure 1B) [1-3]. Two visibly striking examples of such effects can be found with

11   mutations influencing wing development in *Drosophila* and in sexual characteristics of

12   the tail in *C. elegans* (Figure 2A&B). Despite apparent phenotypic similarity in the wild-

13   type state (or in certain environments), there may be considerable segregating genetic

14   variation influencing mutational effects. This so-called *cryptic genetic variation* has been

15   the subject of a number of recent studies with respect to its evolutionary potential [4-11].

16   Simply put, not all "wild-types" are equal.

17        Genetic background effects have been observed in most genetically tractable

18   organisms where isogenic (or pseudo-isogenic) wild-type strains are used, including

19   mice, nematodes, fruit flies, yeast, rice, *Arabidopsis* and bacteria [12-18], [19]. Such effects

20   have also been observed across the spectrum of mutational classes including

21   hypermorphs, neomorphs, hypomorphs, and amorphs [13, 16, 20, 21]. Because they

22   traditionally have been controlled for as "nuisance" variation rather than studied as

23   interesting genetic phenomena in their own right, background-dependent effects are



1    likely to be even more prevalent than current evidence suggests. Here we discuss the

2    importance of considering genetic background effects not only to increase awareness of

3    this issue, but also to argue that by exploiting this variation and integrating knowledge of

4    genetic background, researchers will find increased opportunities for genetic analysis.

5    **Are genetic background effects consequential?**

6    It may be comforting to think that, despite their potential ubiquity, background-

7    dependent effects have only a modest influence on inferences about gene function, but

8    evidence suggests otherwise. Genetic background effects have been implicated in

9    several recent studies, providing explanations for contradictory outcomes and even

10   overturning long accepted results. Several key examples (Boxes 1 and 2) illustrate that

11   a careful consideration of genetic background is crucial for at least two reasons: (i) a

12   failure to control for the genetic background may cause allelic effects at a focal locus to

13   become confounded with variation at other background loci, leading to faulty inferences;

14   and (ii) epistatic interactions between a focal gene and the genetic background may

15   cause different phenotypic outcomes in different genetic backgrounds.

16   Conditional effects may be especially important when considering evolutionary

17   processes, and in particular for evolutionary trajectories. For instance, seemingly

18   phenotypically silent changes in the genetic background of an organism may make later

19   evolution of key innovations accessible. In one example, a long-term experimental

20   evolution line of *Escherichia coli* only evolved a novel trait following certain potentiating

21   mutations [22, 23]. A defining characteristic of *E. coli* is its inability to use citrate as an

22   energy source in aerobic conditions. However, in one lab population of *E. coli*

23   experimentally evolved in a minimal glucose environment (with citrate also present),



1 citrate utilization (Cit+) evolved after about 30,000 generations. Further experiments

2 indicated that at least two potentiating mutations facilitated the origin of this key

3 innovation, and importantly, that it evolved due to an epistatic interaction between the

4 potentiating mutations and the Cit+ mutation, rather than simply an increase in the rate

5 at which Cit+ mutations occur.

6 Similar permissive changes to the genetic background can also facilitate drug or

7 antibiotic resistance—another novel phenotype—by reducing the pleiotropic fitness

8 costs of resistance. For example, the neuraminidase H274Y mutation confers

9 oseltamivir resistance on N1 influenza but compromises viral fitness, and thus had not

10 been commonly observed in natural flu isolates prior to 2007. But in 2007–2008,

11 resistant viruses containing this mutation became prevalent among human seasonal

12 H1N1 isolates. The evolution of oseltamivir resistance was found to be caused by

13 permissive mutations that allowed the virus to tolerate subsequent occurrences of

14 H274Y [24].

15 A number of studies are consistent with the broader idea that the genetic

16 background in which a mutation occurs will influence its evolutionary fate. Several

17 experimental evolution studies show evidence of negative epistasis or even sign

18 epistasis between successive mutations in evolving populations [25-29]. As a result, not all

19 possible evolutionary paths towards an adaptive peak are actually accessible, since

20 some of the paths require a population to traverse a fitness valley. In some cases, the

21 final evolutionary outcome is determined by which mutations have occurred earlier [26, 28].

22 The genetic background may also have more subtle quantitative effects, as



1 demonstrated by one study showing distinct patterns of genetic covariation under

2 mutagenesis in two different genetic backgrounds [30].

3     A key implication of the above observations is that the selection coefficient of an

4 allele can vary depending upon the genetic background in which it is found. Indeed, one

5 study has found evidence for background dependence of selection coefficients on

6 particular alleles of weak to moderate effect [31]. Thus, new models that account for this

7 context-dependent selection will enhance our ability to detect the genomic signature of

8 past selection [32]. Similarly, because the fate of new mutations depends on the genetic

9 background, the repeatability of evolutionary outcomes is likely to be highly dependent

10 upon the genomic context of the ancestral population.

11     These examples also raise questions about the nature of these genetic

12 background variants themselves. For example, what evolutionary forces influence the

13 spread of these background modifier alleles, such as the potentiating mutations in the

14 *E. coli* experiments? One possibility is that without obvious effects on fitness, their

15 spread is dependent on genetic drift. According to this idea of developmental systems

16 drift [33], stochastic forces play a role in determining which regions of "genotype space"

17 are accessible to populations. An alternative is that these potentiating mutations are

18 actually pleiotropic, with effects on other fitness-related traits even in the absence of the

19 focal mutation under investigation. It has been shown that a derived allele influencing

20 vulval phenotypes in *C. elegans* in the presence of sensitizing mutations has a

21 pleiotropic effect on life history traits, which may have helped it spread during laboratory

22 domestication[34]. In another example, evidence is consistent with selection promoting

23 the spread of three permissive mutations that were required for a fourth to enable a



1   phage population to exploit a novel host receptor [35]. In this contrasting view, selection

2   (on unrelated traits) is a central force in making different regions of "genotype space"

3   accessible. These are not mutually exclusive hypotheses, and both chance and

4   selection likely play a role. Nevertheless, this is an under-appreciated aspect to the

5   long-standing debate over the relative importance of selection and drift in determining

6   evolutionary outcomes, and will only be settled with the accumulation of empirical data

7   in diverse organisms.

8   **Should genetic background effects be considered quantitative traits?**

9   Most traits involving morphology, behavior, fitness, and disease are quantitative,

10   displaying continuous variation rather than discrete phenotypes. Such variation is

11   usually a function of many loci of small to moderate phenotypic effects modulated by

12   environmental influences. Nevertheless, for both simplicity and efficiency, many

13   functional genetic analyses still discretize traits, even if these traits could be measured

14   quantitatively, and study the effects of mutant alleles in a tightly controlled manner to aid

15   in inference, even when identifying modifiers (e.g., suppressors and enhancers of a

16   focal mutant allele). Although this approach can substantially simplify the analysis of

17   mutational effects of both the focal allele and its modifiers, it may bias the biological

18   interpretations of allelic effects. For instance, this viewpoint implicitly assumes that

19   background dependence is controlled at least in part by one or more modifiers of major

20   effect.

21   However, an equally plausible alternative is that variation in an allele's effects

22   across two different wild-type genetic backgrounds may be due to variants across many

23   genes. In this case, these genes may interact epistatically, or may have small additive



1  effects (even though these effects are only visible in the presence of the focal mutation).

2  Indeed, the concepts of penetrance and expressivity already provide the necessary

3  framework for this view. For instance, mutations disrupting Ras signaling in *C. elegans*

4  vary quantitatively in the frequencies of different vulval phenotypes induced across

5  different wild-type backgrounds [36]. Likewise, four or more interacting loci are necessary

6  to explain background-dependent variation in the penetrance of many conditionally

7  lethal deletions in *Saccharomyces cerevisiae* [16].

8  Explicitly treating these effects as quantitative rather than discrete traits will

9  allow for a broader set of tools and techniques to be applied to the genetic analysis of

10  context-dependent effects of mutations. Techniques like QTL mapping and association

11  studies can be used to identify polymorphisms associated with variation in expressivity

12  and penetrance (e.g., [1, 2, 34, 37]). The value of this viewpoint is that it is agnostic to the

13  genetic basis of such effects, and with an appropriate density of neutral molecular

14  markers (which will become readily available as whole-genome re-sequencing becomes

15  increasingly affordable), such modifiers can be mapped regardless of their genetic

16  architecture.

17  In particular, "classical" modifier screens involve testing thousands of induced

18  mutants for effects on a focal mutation's penetrance or expressivity. Since any

19  individual induced mutation is unlikely to be a modifier, these studies by necessity look

20  for large effect modifiers. In contrast, moving a focal mutation into a new genetic

21  background nearly always results in subtly different effects. By combining rigorous

22  quantification of these effects with modern genetic mapping approaches, researchers

23  can harness natural genetic variation to detect modifiers with small effects, allowing



1  them to identify a larger and potentially different set of interacting genes [38]. This

2  approach could prove especially useful for geneticists working on a specific genetic

3  pathway or network, particularly when mutagenesis screens have saturated.

4  **The broader context of conditional effects of mutations**

5        A variety of environmental and other factors can alter how a mutant allele

6  influences organismal phenotypes, and the impact of these factors can vary with

7  genetic background. For instance, interactions between developmental temperatures

8  and genetic background influence how a *Distal-less* mutation perturbs leg development

9  in *D. melanogaster* [39]. Larval density and/or nutrition influence both the penetrance and

10  expressivity of antennal duplication of the *obake* mutation[40] and adult foraging behavior

11  for the *rover/sitter* polymorphism [41]. Infection status with *Wolbachia* in *D. melanogaster*

12  can suppress the effects of a mutant *Sxl* allele [42] and influence mutational effects on

13  reproductive success [43]. Even ploidy (which can be considered a form of genetic

14  background) can influence the magnitude of allelic effects [44], as can the genomic

15  location (position effects) of a gene [45]. Indeed, as discussed for genetic background

16  below, not only are the focal mutations' effects context dependent, but so are epistatic

17  interactions between mutations, as illustrated by the host-dependent effects of

18  interacting mutations in Tobacco etch virus [46].

19        Beyond influencing the phenotypic manifestation of large-effect lab-generated

20  mutations, environmental variation frequently modulates the effects of naturally

21  occurring polymorphisms. In *C. elegans*, QTL mapping of life history traits yielded

22  different results at 12°C and 24°C, suggesting distinct loci influence trait variation in

23  different thermal environments[47]. Genome-wide studies imply that these interactions



1     are not rare. For instance, a study mapping variation in transcript levels mirrored this

2     result at the genomic level: a large proportion of expression QTLs (eQTLs) had

3     temperature-specific effects [48]. Likewise, in yeast, a large number of transcripts

4     influenced by eQTLs had environment-specific effects; interestingly, *trans*-acting

5     eQTLs were more likely to have environment-specific effects than cis-acting eQTLs [49].

6         One implication of these results is that it becomes difficult to account for all

7     factors influencing allelic effects. For instance, an investigation of the effects of four

8     natural quantitative trait nucleotides (QTNs) segregating in two yeast strains revealed

9     that trait variation was influenced in a complex way by QTN:QTN interactions which

10     were themselves dependent upon the genetic background and the rearing

11     environment[50]. Thus, what might appear at first to be a two-way QTN:QTN interaction

12     is in reality a higher-order QTN:QTN:background or QTN:QTN:environment interaction.

13     Thus, even when a responsible biologist controls the genetic background and rearing

14     environment of their organism, the scope of their conclusions may be limited to those

15     particular conditions. Of course, many useful discoveries been made by studies using

16     isogenic backgrounds, including the identification of important genes with effects that

17     are apparently consistent across genomic and environmental contexts. However, we

18     still lack enough data to conclude that the genes with "important" roles will generally

19     display similar effects in different situations, and indeed, a failure to control for genetic

20     background may explain conflicting results in several recent studies (Boxes 1 and 2).

21         Such a perspective may also be essential for the future of pharmacogenomics

22     and personalized medicine. For instance, although blocking the EGFR receptor by

23     tyrosine kinase inhibitors is effective against certain forms of cancer, cancers are



1 extremely heterogeneous with variably penetrant mutations in multiple signaling

2 pathways influencing their response to treatments [51-54]. In addition environmental and

3 epigenetic effects influence the occurrence, severity, and drug sensitivity of complex

4 diseases [55]. Studies of such context dependent effects of mutations in model

5 organisms may provide a framework for clinical studies in humans, where

6 investigations of such heterogeneous effects are far more difficult.

7 **Drawing inferences about genetic background effects**

8 When studying the causes and consequences of genetic background generally,

9 and how genetic background effects influence a focal trait specifically, there are a

10 number of issues to consider. One seemingly overlooked issue is having a clear idea of

11 what "trait" is being measured. Consider the influence that genetic background has on

12 the expressivity of the *scalloped*$^{E3}$ (*sd*$^{E3}$) mutation in the *Drosophila* wing (Figure 2). The

13 wings of both wild-type strains (Oregon-R and Samarkand) are qualitatively wild-type,

14 although they differ in size and geometric shape [56]. However, when the *sd*$^{E3}$ mutation is

15 introduced into each of these strains, we observe strong genetic background-dependent

16 effects on wing morphology. As is commonly done in genetic analysis, the measured

17 phenotype (wing morphology) is a proxy for how the mutation perturbs "normal"

18 development. However, adult wing morphology is the result of a complex and dynamic

19 set of developmental events including cell growth, division, death, polarity, and

20 differentiation. The effects of the *sd*$^{E3}$ mutation may influence one or more of these

21 processes. While the differences across genetic backgrounds may be a "strict" genetic

22 background-dependent effect; that is, the mutation perturbs the same developmental

23 processes, but to different degrees in each background. In that case the observed



1   morphological phenotype, and the differences due to genetic background, would reflect

2   the underlying developmental perturbation on a shared set of developmental processes.

3   However, like virtually all other aspects of organismal function, there is considerable

4   variation within and between individuals in these processes. In *Drosophila*, cell

5   proliferation and cell growth vary across wild-type strains [57, 58]. If in one wild-type

6   genetic background cell proliferation was more important for the final size and shape of

7   the wing, while in the other background it was a combination of proliferation and cell

8   growth, then inferences about genetic background effects could be biased. Perhaps the

9   *sd* gene has a greater role in cell proliferation, so perturbing its function disrupts wing

10  development more in the first background than in the second. In this case the observed

11  differences in wing morphology may have less to do with the differential modulation of

12  *sd* function across backgrounds, than with variation in developmental function itself.

13  Although phenomenologically still a background-dependent effect, the developmental

14  and genetic interpretation can be quite different. In this case, for example, there are

15  multiple intermediate traits (Figure 1) underlying the phenotype being measured (wing

16  shape); the mutation's pleiotropic effects (or lack thereof) are responsible for its

17  background dependence.

18      A second example illustrates how background dependence can likewise

19  influence our inferences regarding pleiotropy. A landmark study investigated genetic

20  background effects on mutations that affect the mushroom body and associative

21  odorant learning in *D. melanogaster* [59]. When mutations in 11 genes were crossed from

22  their progenitor background into a Canton-S wild-type background, multiple aspects of

23  the brain qualitatively changed. The authors also examined a wide array of behaviors



associated with brain defects across the original and Canton-S background for an allele of the *mushroom body miniature* gene ($mbm^1$). Although the anatomical phenotypic effects of the mutation were almost completely absent in the Canton-S background, the learning defects remained. This incongruity suggests that the previously inferred causal relationship may have been in part due to the pleiotropic effects of the mutation in the original background, not that the alteration of mushroom body anatomy directly affected learning. Such a disassociation of these supposedly linked phenotypes clearly demonstrates how considering genetic background can help resolve causal links between variation in different traits and lead to a better understanding of pleiotropy.

Finally, the background-dependent phenotypic effect may not reflect the interaction of the background with the lesion per se, but may instead reveal more about other genetic processes, such as the molecular machinery influencing RNAi or the somatic effects of transposable elements on gene expression. Mutations caused by a P-element TE insertion in *D. melanogaster*, for example, are known to show variable penetrance and expressivity because of segregating alleles that suppress P-element activity [60, 61], and these effects may explain the reduced expressivity of mutations when measured in recently wild-caught backgrounds as seen in some studies [62]. Similarly, RNAi-mediated phenotypes might vary in *C. elegans* due to differences in RNAi susceptibility [63] rather than background dependence of specific mutations. Careful interpretation of genetic background effects must therefore also consider whether the effects in question are specific to the focal developmental process or more general properties of a given background.



1  **Where do we go from here: Integrating genetic background effects into genetic**

2  **and evolutionary analyses**

3      Clearly, considering genetic background is essential for researchers seeking a

4  comprehensive understanding of the genotype-phenotype relationship (Figure 1). As

5  others have before, we advocate a research program that controls the genetic

6  background of the focal organism to avoid confounding influences on experimental

7  outcomes. Moreover, we propose that replicating studies across multiple wild-type

8  genetic backgrounds will not only help biologists clearly establish the generality of their

9  findings, but will also help identify larger sets of interacting genes, particularly genes

10  with small effects. Although this approach requires the investment of time and

11  resources, it will provide a less biased view of genetic networks and enable more

12  precise predictive models for today's complex research areas (e.g., personalized

13  medicine). Practical measures can be taken to balance the tradeoff between resource

14  investments and generality of conclusions (Box 3). For instance, in more tractable

15  organisms such as yeast, transgenics could be made in multiple wild backgrounds.

16  When time is an issue, using chromosome substitution (e.g., with balancers as in

17  *Drosophila*) rather than introgression by backcrossing can provide to a first

18  approximation, the background dependence of a mutation's effects (and provides the

19  added benefit of simultaneously mapping any background modifiers to a specific

20  chromosome).

21      For evolutionary geneticists, investigating the background dependence of an

22  allele's effects can lead to an improved understanding of how selection acts on that

23  allele [17, 44]. As previously mentioned, [50] the effects of four natural QTNs between two



1 yeast strains have been investigated in detail. Although the QTN effects were

2 consistent in direction across backgrounds and environments, their magnitudes, and

3 those of the QTN:QTN interactions, varied, meaning that selection on them will also

4 vary. Likewise, interest in the various forces that can influence the selection coefficient

5 on an allele, such as sexual selection, has also surged [64-70]. However, the basic

6 framing of this question depends on the genetic context, and allelic effects (and thus

7 selection coefficients) likely vary across backgrounds. How this variability influences

8 the evolutionary dynamics of allele frequencies thus remains an important open

9 question.

10 Another important consequence of background dependence on evolution is that,

11 because an allele's effects depend on the genetic milieu, the genetic background can

12 limit the types of phenotypes that are evolutionarily or mutationally accessible (e.g., [28,

13 71]). An outcome (e.g., parallel molecular evolution) in an experimental evolution study

14 (particularly one beginning with an isogenic strain, as in many microbial studies) may

15 be repeatable only in that genetic background; repeating the study with different

16 genetic backgrounds may yield alternative outcomes, with the potential to change our

17 views on how prevalent convergence is at the genetic level. We therefore believe that

18 efforts should be made to initialize experimental evolution populations with multiple

19 backgrounds, in addition to multiple replicates from a single isogenic ancestor.

20 Although the influences of genetic background and the environment have been

21 recognized since the early days of genetic analysis—and indeed, many conclusions

22 based on studies in isogenic lines have provided valuable generalizable insights—their

23 effects on mutational interactions (epistasis) were assumed to be negligible. But as



demonstrated in the examples above [1, 16 17, 46], if genetic interactions as inferred from mutational studies are influenced by genetic background, then we are ignoring an implicit fact that epistatic interactions are themselves background dependent. Thus the choice of the genetic background used in an interaction or sensitization screen can significantly alter its outcome, including the number of modifiers identified as well as the direction and magnitude of their effects. Indeed, mapping of the background-dependent effects may yield additional modifiers, painting a more complete picture of the genetic network being studied. The topologies of the genetic networks inferred from these interaction studies may in fact turn out to be more variable than currently appreciated. For those who aim to chart the genotype-phenotype map—whether to make predictions about health-related traits or the outcome of natural selection—knowing the full topology of these genetic networks is essential; including information on variable interactions will improve predictions of phenotypes from genomic data.

In addition, a number of questions about the nature of genetic background effects themselves remain underexplored. At the most basic level, though genetic background effects can clearly confound genetic analyses, we lack sufficient data to generalize how often this occurs and in what situations the problem is most severe. For instance, are mutant alleles with small effects on organismal phenotypes more subject to modulation by genetic background than large-effect mutations? We also know little about the genetic architecture of genetic background effects, such as the number and effect size distribution of the causal background polymorphisms. In addition, a better understanding of how pleiotropy can vary with genetic background is essential for understanding relationships between traits. These questions can only be answered by



1 additional empirical studies, e.g., surveys and mapping studies of genetic background

2 effects involving different allele types and a range of organisms.

3 We understand that performing a complex experiment involving multiple genetic

4 backgrounds and/or environments is difficult and complicates interpretations. But then

5 any conclusions drawn from studies in a single background must be recognized to

6 have a limited scope with respect to allelic effects, gene structure-function

7 relationships, pleiotropy, and epistasis. Despite the additional workload, the payoff for

8 performing such studies across multiple wild-type backgrounds therefore has the

9 potential to profoundly transform our understanding of genetics and the genotype-

10 phenotype relationship.




12 Acknowledgements: We thank Greg Gibson, Ellen Larsen and members of the

13 Dworkin lab for insightful discussions. We would like to thank the two anonymous

14 reviewers and the editor (Dr. Rhiannon Macrae) for suggestions that have significantly

15 improved this manuscript. This work was supported by the National Science

16 Foundation under MCB-0922344 and NIH grant 1R01GM094424–01 (to ID). The

17 authors have no conflict of interests to declare.












**Box 1: Genetic inferences about longevity and genetic background effects**

Contradictory results across studies may be due to differences in or a lack of controlling for wild-type genetic backgrounds.  We discuss two particular examples on the genetics of aging, which could have significant clinical and economic impact. The *I'm not dead yet* (*Indy*) gene of *Drosophila* was initially implicated in extending lifespan: flies heterozygous for loss-of-function alleles of *Indy* were reported to have increased life span in the Canton-S wild-type background [73]. However, when the mutations were later outcrossed into a large natural population or backcrossed into additional isogenic wild-type strains, most of the mutational effects disappeared[74]. Instead, additional mutations independent of *Indy* seemed responsible for increasing lifespan. Thus many of *Indy*'s previously reported effects likely represent interactions between *Indy* mutations and genetic background (including inbreeding) [75], in addition to *Indy*-independent mutations and environmental effects [74, 76].  Despite this, these mutants were used in recent studies [77, 78], resulting in disagreements on interpretation and a discussion of which isogenic "wild-type" backgrounds the longevity effects are apparent in[79, 80] (although no discussion of why they differ).

The role of the *sir2* gene in longevity has also been reconsidered because of genetic background effects. Despite years of research into the role of the *sir* genes on lifespan [81], two high-profile papers failed to replicate key results[82, 83]. Instead, the extended lifespan of transgenic *C. elegans* was the result of a secondary mutation, not the *sir2-2.1* transgene itself. In *Drosophila*, backcrossing flies to the appropriate wild-type strain



eliminated the increased lifespan associated with overexpression of sir-2.1 [82]. The

implications of these findings have been extensively debated [84-86].  As with the example

above, it is not clear whether these discrepancies are due to true background-

dependent effects (i.e., different backgrounds respond to the transgene differently), or

artifacts from a failure to control the genetic background (i.e., genetic background is

confounded with the focal mutation). Indeed, the wild-type *Drosophila* strain that

suppressed the lifespan-increasing effects of *sir-2.1* overexpression was Dahomey, in

which *Indy*'s effects also disappeared [74]. One plausible  (but untested) explanation is

that Dahomey is suppressive of mutations influencing longevity. If so, investigating the

effects of these mutations in other isogenic wild-type backgrounds may yield different

results [80].

These examples raise two important issues. First, is it ever sensible to perform genetic

experiments in only a single wild-type background? Second, how do you ensure that

two genetic backgrounds with the same name are in fact genetically similar or identical

(given that new mutations accumulate in lab cultures)? We discuss these problems

further in Box 3.



**Box 2: Genetic Background effects and evolutionary inferences**

One of the early experiments to use gene replacement in *Drosophila melanogaster* investigated the influence of naturally occurring polymorphisms in the *desat2* (*dz*) gene [87], thought to be involved in the synthesis of contact pheromones (cuticular hydrocarbons). Molecular evolution studies suggested *desat2* was under divergent selection in two populations of *D. melanogaster*, with a potential role in premating isolation between flies from Zimbabwe and the cosmopolitan "population". Greenberg et al [87], integrated both the cosmopolitan $dz^M$ allele (likely loss of function) and the $dz^2$ allele found in Africa and the Caribbean into a common genetic background for comparison. There was no evidence that variation in *dz* mediates mate discrimination, but the data suggested that *dz* influenced other ecologically relevant traits. However, one of the co-authors of the original study later reported that attempts to replicate it failed [88, 89]. In a reply, Greenberg et al [90] suggested that no attempt was made to control for genetic background in the re-analysis. A similar pattern emerged in the analysis of the role of the *tan* locus's contribution to pigmentation differences between two closely related *Drosophila* species (for more details, see [87-89]). In both examples, the exact contribution of genetic background was never clearly established. The differences might have been caused by epistatic interactions between the focal alleles and the different genetic backgrounds. Alternatively, the focal alleles may have become confounded with additional background variants influencing the traits, resulting in a spurious correlation between the phenotypes and the focal alleles. In the former case, any inferences about the evolutionary processes leading to the fixation of these alleles would need to account



1    for the epistatic interactions between each allele and the genetic background.



1     Box 3. Considerations for research programs incorporating genetic background.

2     **1. *How many genetic backgrounds is enough?*** A balance between practical

3     consideration, research goal, and generality of conclusions needs to be struck. If

4     the goal is to understand the distribution of genetic background effects for a small

5     number of mutations, then tens to dozens (flies, *C. elegans*, *Arabidopsis*) or more

6     (yeast, bacteria) may be suitable.  If the goal is instead to broaden a specific set

7     of genetic inferences (structure-function, modifier screens, epistasis), then only a

8     few genetic backgrounds may be practical for most organisms. If replacing the

9     entire genetic background is impractical, efforts should be made to at least

10    perform preliminary crosses, such as balancer-mediated replacement of

11    individual chromosomes (mice, *Drosophila*).

12    2. ***Isogenic (inbred) strains, outbred populations, or somewhere in between?***

13    Isogenic inbred wild-type strains may not always be optimal for particular

14    research questions. Traits closely tied to fitness are susceptible to inbreeding

15    depression in some organisms (*Drosophila*, mice), but less so in others

16    (*Arabidopsis* and *C. elegans*). Inbreeding creates additional genetic stress,

17    independent of the focal mutation, influencing traits like longevity [75]. Yet crossing

18    mutations into outbred populations may be problematic, making it difficult to

19    partition genetic effects, and "average" phenotypes may be biased. If the

20    mutation is lethal with certain combinations of naturally occurring alleles in the

21    base population, then some combinations of alleles may be unobserved.  Even

22    when a measure such as the selection coefficient for an allele is examined, an

23    outbred population may not be averaging the fitness cost of an allele *per se*, as



variants present in the population may be under selection to compensate for the focal allele.

If measuring mutational effects in an inbred line is problematic, crosses between inbred strains can generate "clonal" $F_1$ individuals, ameliorating inbreeding (reciprocal crosses may be necessary if maternal effects are suspected). This will require introgression of the focal allele into multiple inbred lines, followed by experimental production of $F_1$s.

3. ***How do you know your background is what you think it is?*** Certain sub-fields commonly use the same apparent background, at least in name. Setting aside the non-trivial issue of contamination of wild-type stocks, there are several issues to consider.

Researchers often introduce visible markers into given backgrounds, but this may also introduce linked genomic fragments. Moreover, "copies" of strains kept in separate labs will accumulate new, independent mutations, or fixation of different (residual) segregating alleles, especially when maintained at low population sizes[91]. Thus a combination of fresh inbreeding, and genotyping by re-sequencing or other methods, may be necessary to confirm the identity of a particular genetic background.

4. ***How do you get your mutation into each wild-type strain?*** Introducing mutations into multiple backgrounds is often the greatest barrier to this work. In some organisms (*Drosophila*, mice), introgression of the allele into multiple



backgrounds occurs by backcrossing, which is labor-intensive, requiring months or years for sufficient introgression. This technique also results in introgression of genomic regions linked to the focal allele, with the size of the introgressed fragment varying across backgrounds (potentially requiring multiple independent replicates for each background). Although this technique will remain an essential tool for the near future, transgenic techniques including homologous gene replacement and gene knockouts [92] in multiple backgrounds will hopefully become widely available.

Additionally, transgenic inserts that knock down gene function using RNAi are becoming widely available [93]{ and can be inserted into the same genomic location (minimizing positional effects). Although this may introduce additional complications (e.g., genetic background influencing RNAi machinery, not the focal gene; off-target effects [94]; RNAi machinery itself influencing phenotype [18]), it may be more feasible to generate these in multiple independent genetic backgrounds[72].



1   **Glossary**



3   **Penetrance-** The proportion of individuals in a sample with a particular genotype

4   expressing the "expected" phenotype.



6   **Expressivity-** The extent to which a mutant genotype is phenotypically expressed in an

7   organism. Often, mutations may display variable expressivity; i.e., multiple individuals

8   carrying the same mutation may vary for the phenotypes induced by the mutation.



10   **Cryptic genetic variation-** Genetic variation present in a population that is not

11   phenotypically expressed under benign or ambient conditions, but which may be visible

12   upon genetic or environmental perturbations.



14   **eQTL-** A sequence polymorphism in the genome associated with variation in gene

15   expression.



17   **Wild-type-** The "average" phenotype, often assumed to be the "normal" phenotype,

18   found in natural populations and/or any subpopulation or inbred lines derived from such

19   a population. The genotypes producing such a phenotype are often considered to be

20   wild-type genotypes.



22   **Genetic background-** An organism's entire genetic and genomic context; the complete

23   genotype of an organism across all loci.



**Isogenic-** Having identical (or nearly identical) genotypes.

**Line/strain-** A distinct interbreeding population, usually maintained in the lab, and which is isolated from other such populations, often generated by inbreeding.

**Potentiating/permissive mutations-** Mutations that are required to occur first in order for subsequent mutations to be expressed.

**Introgression –** The introduction of an allele or alleles from one population into another by repeated backcrossing.

**Amorph/hypermorph/hypomorph/neomorph-** Mutant alleles exhibiting no activity, increased activity or expression, reduced activity or expression, and some novel activity, respectively.



1  **Figure Legends**



3  **Figure 1.** Genetic background effects can be conceptualized in the framework of a

4  genotype-phenotype map [95-98]. (A) A wild-type genotype at a particular locus results in a

5  wild-type final phenotype (gray circle), even though there may be variation in

6  intermediate (e.g., gene expression) and "final" phenotypes among different genetic

7  backgrounds (or in different environments). Each color represents a distinct genotype or

8  strain. (B) However, when a particular gene is mutated, intermediate variation among

9  different genetic backgrounds may be expressed in the form of distinct final mutant

10  phenotypes (with some possibly overlapping with the range of wild-type phenotypes

11  (gray circle), and others being distinct). The general increase in variation between

12  backgrounds under the mutational perturbation (i.e. the "cryptic genetic variation") is

13  depicted by the broader distributions of final phenotypes in panel B.  Finally, while this

14  and many other representations of the G-P map represent the genotypic space as a

15  simple projection (much like the intermediate "phenotypic" spaces), it is important to

16  remember that the different genotypic spaces interact as well (i.e., the phenotypic

17  outcomes depend on the position in both genotypic spaces, not simply the position in

18  the "lowest" genotypic space).



20  **Figure 2.** Induced mutations often have qualitatively or quantitatively variable effects on

21  organismal phenotypes in different genetic backgrounds and in different environments.

22  These effects can range from mild (in some cases, perhaps even resulting in

23  phenotypes that are indistinguishable from the wild-type) to severe. (A) The *scalloped$^{E3}$*



allele has qualitatively distinct effects on wing morphology in two commonly used wild-type strains of *Drosophila melanogaster*, despite the wild-type wings being qualitatively similar across these backgrounds. These background effects extend to include epistatic interactions between *sd* and other loci [1]. (B) The effects of the *tra-2*(*ar221*); *xol-1*(*y9*) genotype on sexual differentiation in the tail of *Caenorhabditis elegans* vary quantitatively with both rearing temperature and wild-type genetic background [2]. The effects of genetic background are most apparent at intermediate temperatures.


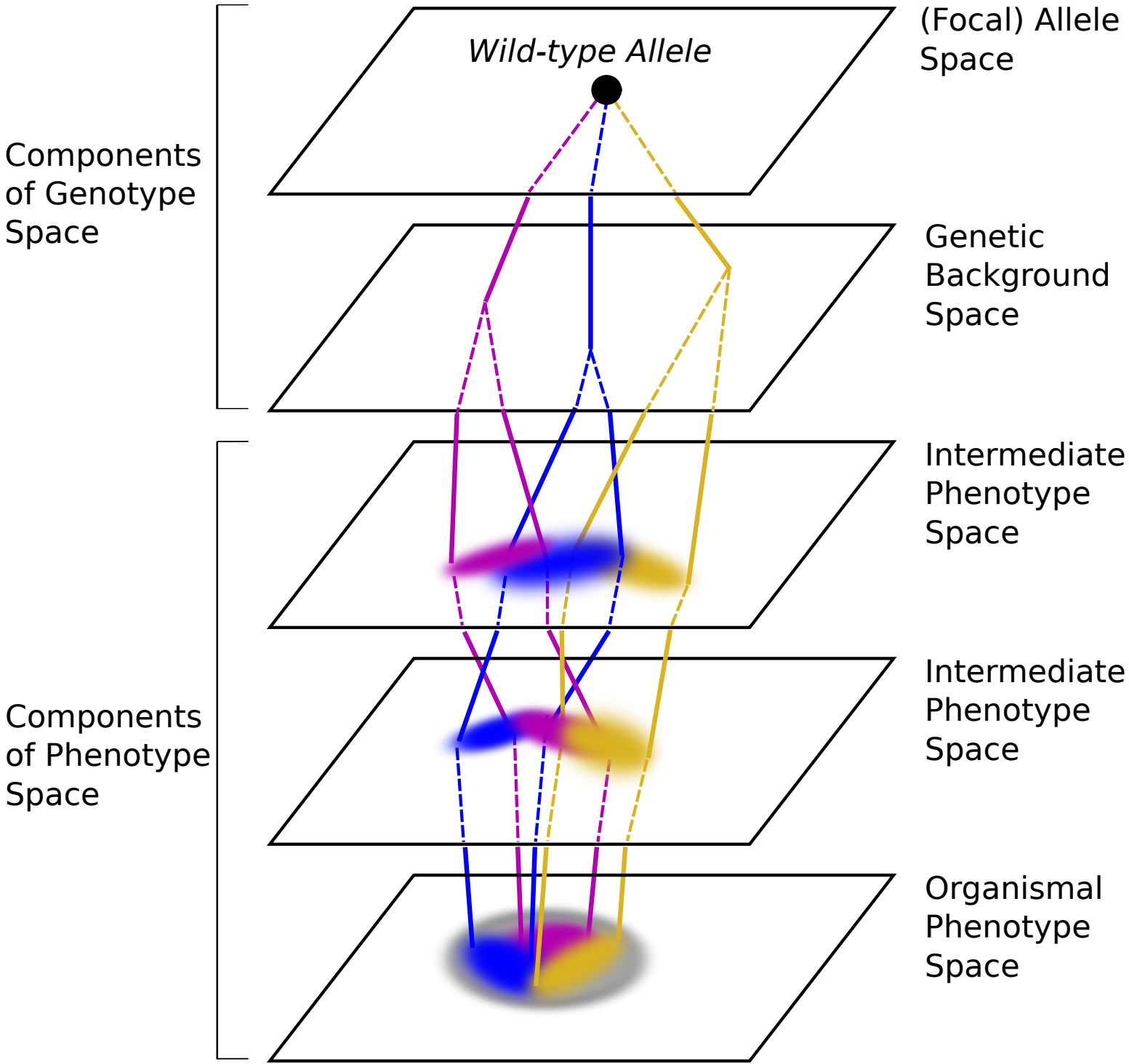

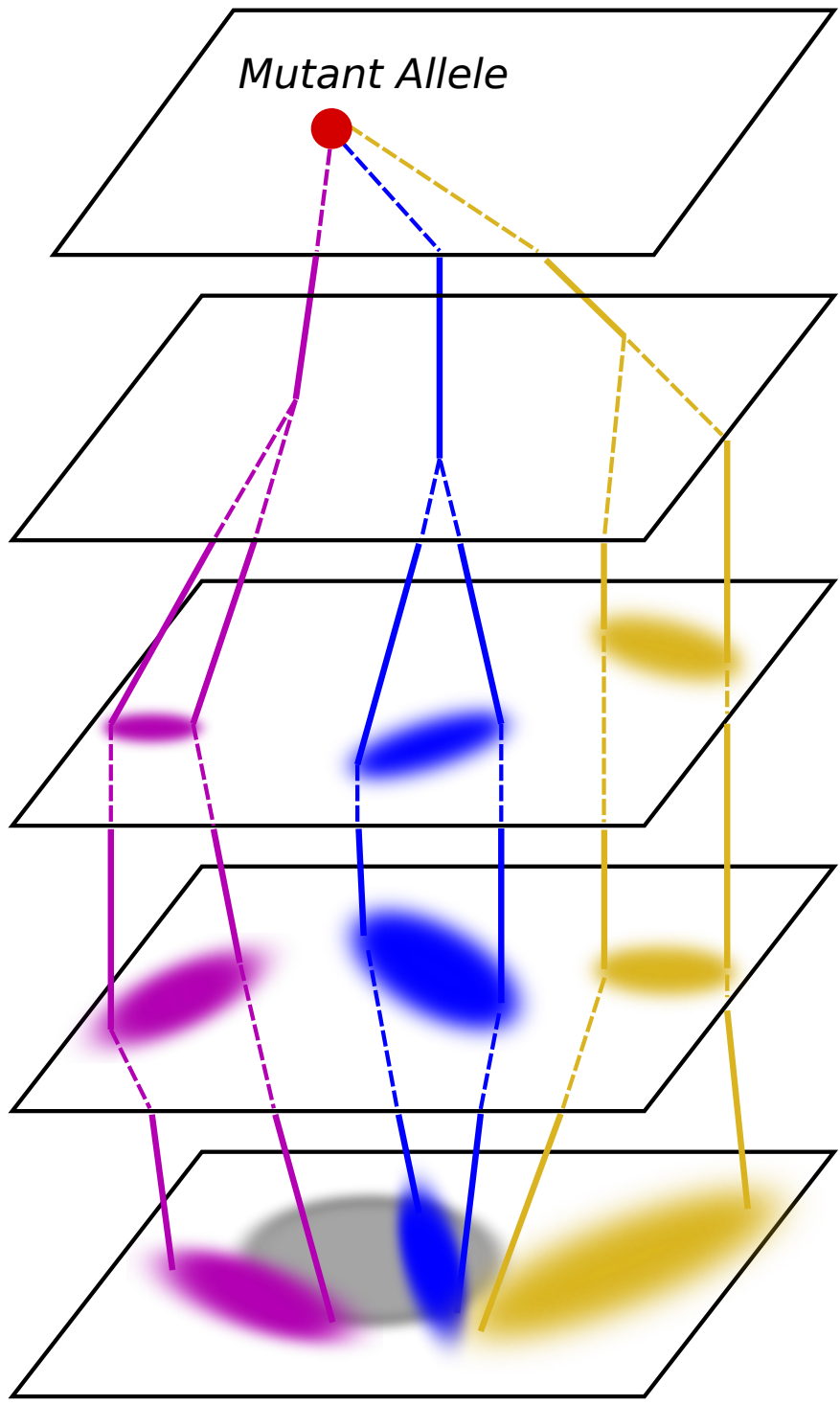

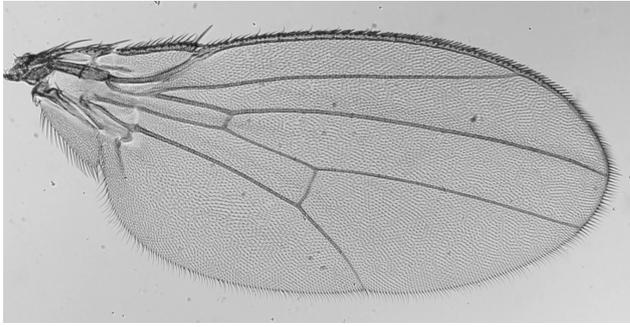

**Wild type**

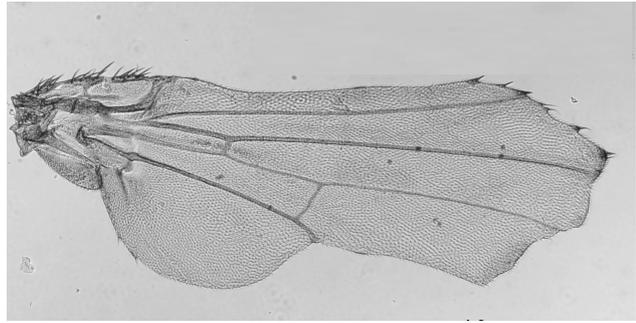

**Samarkand $sd^{E3}$**

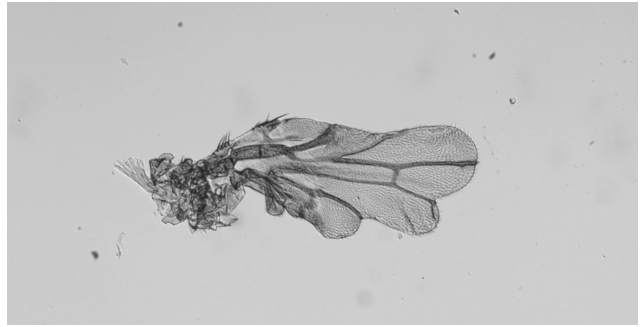

**Oregon-R $sd^{E3}$**

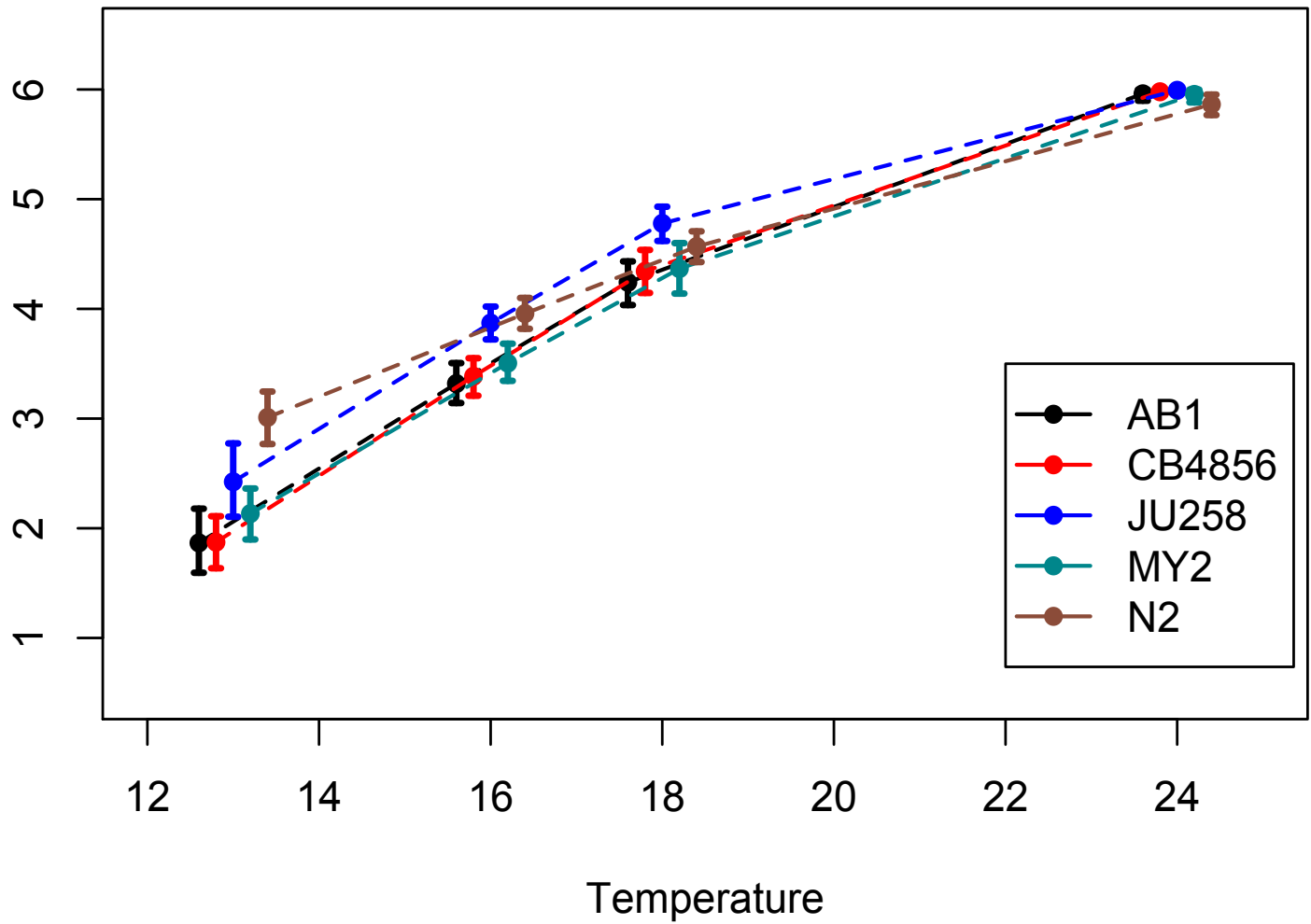
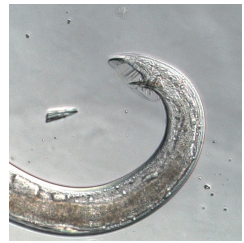
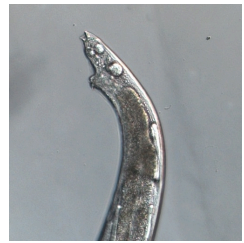
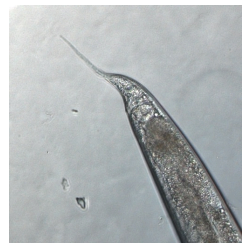